\documentclass[%
 reprint,
superscriptaddress,
 amsmath,amssymb,
 aps,prl,floatfix
]{revtex4-2}

\usepackage{graphicx}% Include figure files
\usepackage{dcolumn}% Align table columns on decimal point
\usepackage{bm}% bold math
\usepackage[english]{babel}
\usepackage{soul}

\usepackage{hyperref}
\hypersetup{pdfstartview={FitH},pdfpagemode={UseNone},
            colorlinks,linkcolor=blue, citecolor=blue, urlcolor=blue,
            bookmarksopen=true, pdfnewwindow=true}
\usepackage[all]{hypcap}

\usepackage{braket}

\begin{document}

\preprint{APS/123-QED}

\newcommand{\ANU}{ARC Centre of Excellence for Transformative Meta-Optical Systems (TMOS), Department of Electronic Materials Engineering, Research School of Physics, The Australian National University, Canberra, ACT 2601, Australia}
\newcommand{\JENA}{Institute of Applied Physics, Abbe Center of Photonics, Friedrich Schiller University Jena, Albert-Einstein-Straße 6, Jena 07745, Germany}
\newcommand{\OULU}{Opto-Electronics and Measurement Techniques, Advanced Electronics Centre, Faculty of Information Technology and Electrical Engineering, University of Oulu, Oulu, FI-90014, Finland}
\newcommand{\SZU}{Institute of Quantum Precision Measurement, State Key Laboratory of Radio Frequency Heterogeneous Integration, College of Physics and Optoelectronic Engineering, Shenzhen University, Shenzhen 518060, P. R. China}
\newcommand{\ASTON}{Aston Institute of Photonic Technologies, Aston University, Birmingham, B4 7ET, UK}
\newcommand{\FRAUNHOFER}{Fraunhofer Institute for Applied Optics and Precision Engineering IOF, Albert-Einstein-Straße 7, Jena 07745, Germany}

\title{Quadratic Quantum Polarimetry with Entangled Photon Pairs}% Force line breaks with \\
%Two-Photon Quantum Polarimetry with Entanglement for Scattering Media

\author{Jinliang Ren}
\affiliation{\ANU}

\author{Vira Besaga}
\email{vira.besaga@uni-jena.de}
\affiliation{\JENA}

\author{Ivan Lopushenko}
\affiliation{\OULU}

\author{Jinyong Ma}
\affiliation{\ANU}
\affiliation{\SZU}

\author{Alexander~Bykov}
\email{alexander.bykov@oulu.fi}
\affiliation{\OULU}

\author{Igor Meglinski}
\affiliation{\ASTON}

\author{Frank Setzpfandt}
%\email{f.setzpfandt@uni-jena.de}
\affiliation{\JENA}
\affiliation{\FRAUNHOFER}

\author{Andrey A. Sukhorukov}
\email{andrey.sukhorukov@anu.edu.au}
\affiliation{\ANU}

\date{\today}% It is always \today, today,

\begin{abstract}
\begin{description}
\item[Abstract]
Conventional polarimetry, including schemes leveraging entangled light, characterizes optical samples through linear transformations of polarization states. 
%However, higher-order polarization correlations, which carry additional information about complex scattering and depolarization processes, remain inaccessible. 
We introduce a two-photon probing approach in which both photons of an entangled pair interact with the same depolarizing medium simultaneously. In this regime, the transformation of the two-photon polarization correlations becomes quadratic in the Mueller matrix, enabling access to second-order polarization information beyond conventional polarimetry. We develop a theoretical framework linking the Mueller matrix to the evolution of the two-photon polarization correlation tensor and show that depolarization induces quadratic degradation of entanglement and state purity. Experiments using polarization-entangled photon pairs transmitted through controlled scattering media confirm the predicted response and reveal enhanced sensitivity to polarization scrambling compared with single-photon probing. These results establish two-photon probing as a higher-order quantum polarimetric modality for characterizing polarization channels.\\

\item[keyword]
Quantum entanglement, Quantum polarimetry, Stokes tensor congruence transformation, Higher-order polarization correlations, Stabilizer-induced non-uniqueness
\end{description}
\end{abstract}

\maketitle

%\tableofcontents

\begin{figure}
    \centering
    \includegraphics[width=1\linewidth]{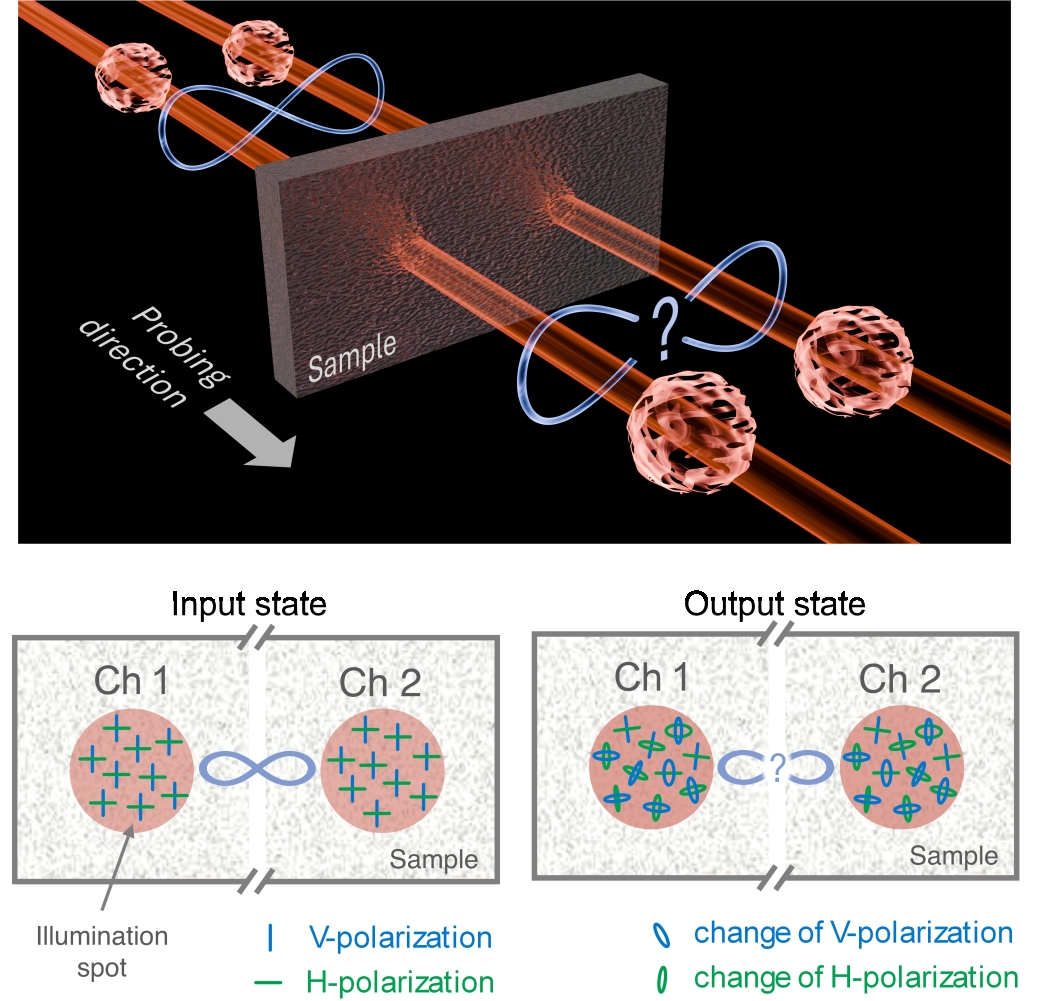}
    \caption{Concept of quantum two-photon probing polarimetry (TPP). A pair of polarization-entangled photons is directed into two spatially separated illumination channels (Ch 1 and Ch 2) of a scattering sample. While the photons initially share well-defined polarization correlations, multiple scattering within the medium modifies the local polarization of each photon and may degrade their mutual coherence. %By analyzing the polarization state in each output channel, TPP enables characterization of polarization transformation and entanglement degradation with enhanced sensitivity.
    }
    \label{fig:Concept}
\end{figure}
\emph{Introduction}. Polarization transformations of optical fields are traditionally characterized using Stokes vectors and Mueller matrices, which describe linear mappings between input and output polarization states~\cite{qi2018,cloude1990SPIE}. This framework forms the foundation of classical polarimetry and is widely used to study anisotropic, depolarizing and scattering media~\cite{azzam1977ellipsometry,He:2021-194:LSA}. Whereas conventional polarimetric measurements access only first-order polarization moments of the optical field, there is a fundamental and practical interest in exploring distinct approaches that can deliver enhanced sensitivity or access to additional information associated with complex light scattering.
%therefore provide an incomplete description of polarization correlations in complex optical channels.

Quantum light offers new opportunities for probing polarization properties beyond the classical Stokes formalism. In particular, entangled photon pairs exhibit non-classical polarization correlations that have been exploited in quantum metrology and sensing~\cite{Goldberg:2021-1:ADOP, Giovannetti:2004-1330:SCI}. This has led to the development of quantum polarimetry, where polarization properties of a sample are probed using quantum states of light to access correlations beyond classical %\hl{\st{first-order moments}
~\cite{Zhang:2024-eadk1495:SCA,Pedram:2024-2400059:ADQ,Goldberg:2020-23038:PRR,Zhang:25:Cells}. In most existing quantum polarimetric schemes, however, only one photon interacts with the sample while the second photon serves as a reference~\cite{Elsevier2022Goldberg,Xie:2025:npjQuantum,Wolfgramm:2013:NatPho}. As a result, the measured transformation remains effectively equivalent to single-photon polarimetry and does not fundamentally alter the structure of the polarization channel being probed~\cite{Safadi:2023-562:NPHYS,Lib:2022-986:NPHYS}.

Here we introduce a fundamentally different configuration in which both photons of an entangled pair simultaneously propagate through the same polarization channel. We show that this two-photon probing regime transforms the polarization measurement from a linear Mueller mapping into a quadratic channel acting on the two-photon polarization correlation tensor. When a medium described by Mueller matrix $M$ acts identically on two photons, the two-photon polarization correlation tensor $K$ evolves according to $K' = M KM^T$ which is quadratic in $M$, revealing a second-order polarization structure not directly accessed in standard first-order Stokes-Mueller polarimetry. This quadratic mapping implies that two-photon probing directly accesses second-order polarization correlations and fundamentally changes the response of polarization observables to depolarization.

We develop a theoretical framework linking Mueller matrices, quantum channels, and two-photon polarization correlations, and derive analytical predictions for the evolution of entanglement and purity under depolarizing channels. The theory predicts a quadratic amplification of depolarization effects compared with single-photon probing. We experimentally verify these predictions using polarization-entangled photon pairs transmitted through controlled scattering media. The observed degradation of concurrence and purity confirms the predicted quadratic sensitivity and demonstrates that two-photon probing provides access to second-order polarization information. Our results establish quantum two-photon polarimetry as a higher-order measurement paradigm for characterizing polarization channels.

\emph{Theory.} Classical polarization transformations in optical media are commonly described using the Stokes–Mueller formalism. In this framework the polarization state of light is represented by the Stokes vector $S = (S_0, S_1, S_2, S_3)^T$, and its evolution through an optical system is governed by a Mueller matrix $M$, $S' = MS$~\cite{chipman2018polarized,chipman1995mueller}. This linear relation forms the basis of conventional polarimetry and has been widely used to characterize birefringence, diattenuation, and depolarization in complex optical media. Physically, this formalism captures transformation of the first-order polarization moments of the electromagnetic field. In quantum optics the polarization state of a single photon can be described by a density matrix~\cite{Goldberg:2021-1:ADOP,Lung:2024-1060:ACSP,Fano:1949-859:JOS},
\begin{equation}
    \rho = \frac{1}{2}\left( \mathbb{I} +\sum_{i=1}^3S_i\sigma_i\right), \quad i=1,2,3,
\end{equation}
\noindent where $\sigma_i$ denote the Pauli matrices and $\mathbb{I}$ is the identity matrix. The interaction of a photon with a scattering medium can then be modeled as a completely positive trace preserving (CPTP) quantum channel by averaging over the ensemble of $N$ realizations~\cite{Choi1975,Jamiolkowski1972},

\begin{equation}\label{eq:CPTP}
    \rho_{\text{out}} = \frac{1}{N}\sum_{k=1}^N U_k\rho_{\text{in}}U_k^\dagger,
\end{equation}

\noindent where the Kraus operators $U_k$ encode the polarization transformation of the $k^{\text{th}}$ realization and satisfy $\sum_k U^\dagger_kU_k=\mathbb{I}$. For a single photon traversing the medium, $U_k=J_k$ is the Jones matrix. In the case when only one photon out of the entangled pair (photon 1) interacts with the sample and its partner (photon 2) acts as a reference, $U_k = J^{(1)}_k\otimes\mathbb{I}^{(2)}$.%, where $\mathbb{I}^{(2)}$ identity matrix on photon 2.

While this description fully captures the behavior of single-photon polarization states, it does not account for higher-order polarization correlations that arise in multi-photon quantum states. We now consider a configuration in which both photons of a polarization-entangled pair propagate through the same medium. If each photon experiences statistically identical but microscopically independent polarization channel, the joint transformation of the two-photon density matrix becomes
\begin{equation}
    \rho_{\text{out}} = \frac{\sum_{k_1,k_2=1}^{N_{k_1},N_{k_2}} \left(J_{k_1}^{(1)}\otimes J_{k_2}^{(2)}\right)\rho_{\text{in}} \left(J_{k_1}^{(1)}\otimes J_{k_2}^{(2)}\right)^\dagger}{N_{k_1}N_{k_2}},
\end{equation}
where the indices $k_1$ and $k_2$ label independent microscopic scattering realizations of the partner photons labeled by the corresponding superscripts. %photon pairs.
The polarization correlations of the two-photon state can be described by the correlation tensor and then the Mueller matrix can be expressed using the CPTP map,
\begin{equation}\label{eq:KMU_map}
\begin{aligned}
      &K_{ij} = \text{Tr}\left[\rho(\sigma_i\otimes\sigma_j)\right],\\
      &M_{ij} = \frac{1}{2}\text{Tr}\left[\sigma_i\frac{1}{N_\mu}\sum_{\mu=1}^{N_\mu } J_\mu\sigma_jJ_\mu^\dagger\right],
\end{aligned}
\end{equation}
\noindent which characterizes the joint expectation values of the Pauli operators for the two photons. Using the relation between Pauli operators and the Mueller matrix representation of statistically identical polarization channels, one finds that the correlation tensor evolves according to (see Supplementary Material, Secs.~S1 and S2),
\begin{equation}\label{eq:congurence}
    K_{\text{out}} = MK_{\text{in}}M^T.
\end{equation}

\noindent This form makes explicit that TPP probes second-order polarization correlations through a bilinear action of the Mueller matrix. This quadratic mapping %of two-photon probing 
represents a fundamental difference with respect to single-photon probing. Whereas conventional quantum one-photon polarimetry measures the linear transformation of first-order polarization moments, the two-photon configuration probes second-order polarization correlations encoded in the joint quantum state. Since $M$ is the effective Mueller matrix, this relation applies independently of the microscopic origin of depolarization, including multiple scattering, dispersion, or random birefringence.

\emph{Sensitivity to depolarization.} The physical consequences of this quadratic transformation can be illustrated using a simple depolarizing channel described by the Mueller matrix, $M = \text{diag}(1,m_{11},m_{22},m_{33})$, where $0\leq m_{ii}\leq1$ characterizes the degree of polarization preservation. 

To compare quantum one- and two-photon polarimetry, we consider a Bell state input~\cite{Mishchenko2002,Kwiat:1995-4337:PRL}, $\ket{\Psi^+} = 1/\sqrt{2}(\ket{HV}+\ket{VH})$, which produces a diagonal correlation Stokes tensor $K_{\text{in}}=\text{diag}(1,-1,1,1)$. The formalism imposes no specific requirement on the input state, while Bell-type states are used here as an example.
%for the maximum correlation contrast.
%and counting statistics.

For quantum one-photon polarimetry (OPP), only one photon interacts with the sample and another acts as reference, so the polarization channel acts linearly on one subsystem. Then the output purity $\gamma=\text{Tr}(\rho^2)$ becomes
\begin{equation}
    \gamma_{\text{OPP}} = \frac{1}{4}\left[1+m_{11}^2+m_{22}^2+m_{33}^2\right].
\end{equation}

For two-photon polarimetry (TPP), both photons traverse the same channel, and the output purity becomes
\begin{equation}
    \gamma_{\text{TPP}} = \frac{1}{4}\left[1+m_{11}^4+m_{22}^4+m_{33}^4\right].
\end{equation}

For this simplified model, in OPP, the dependence on depolarization is quadratic in the Mueller matrix elements, whereas in TPP it becomes quartic. The origin of this difference is precisely the quadratic channel action given in Eq.~(\ref{eq:congurence}). The input state can be generalized to arbitrary state to observe the quartic purity. The same argument applies to concurrence and related correlation-based observables, once the channel acts on both photons, the depolarization factors enter multiplicatively and amplify the loss of polarization correlations.

\begin{figure}
    \centering
    \includegraphics[width=1\linewidth]{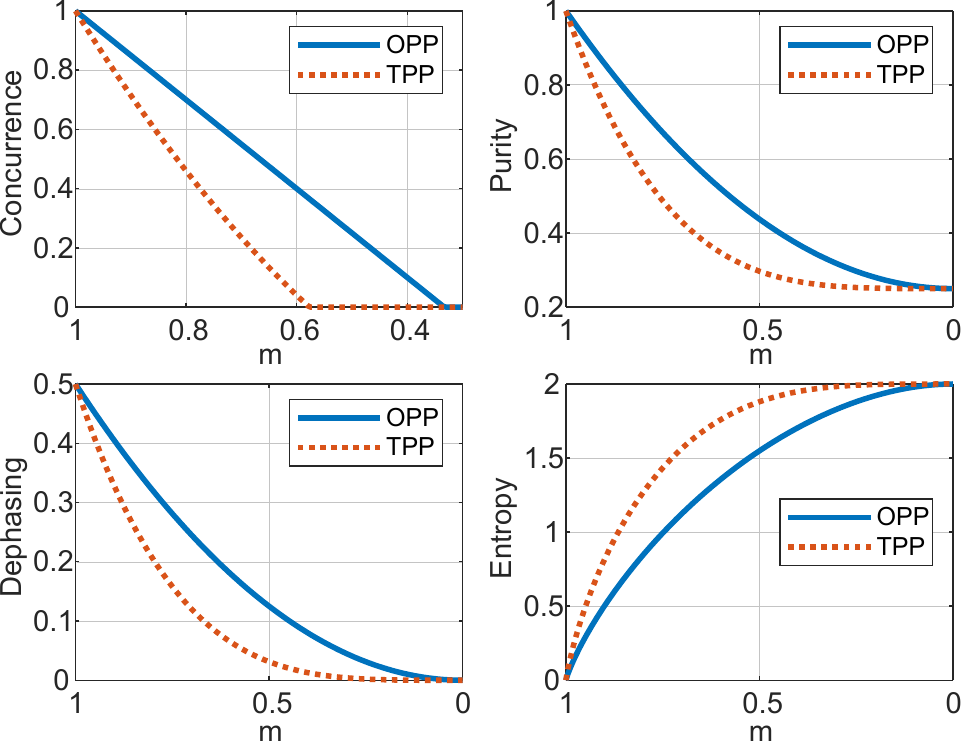}
    \caption{Theoretical comparison between one-photon polarimetry (OPP) and two-photon polarimetry (TPP) for a diagonal depolarizing channel $M=\text{diag}(1,m,m,m)$. Shown are the concurrence, purity, dephasing strength, and von Neumann entropy of the output state as functions of $m$. The TPP response exhibits a stronger, nonlinear dependence on $m$ leading to accelerated degradation of coherence and entanglement.}
    \label{fig:CPDE}
\end{figure}

This also clarifies the enhanced sensitivity. For clarity of expressions, we consider the isotropic case, $m=m_{11}=m_{22}=m_{33}$, with $\gamma_{\text{OPP}} = (1+3m^2)/4$ and $\gamma_{\text{TPP}} = (1+3m^4)/4$. If we parametrize the scattering strength of a medium under test by an effective thickness $\eta$, with $m:=m(\eta)$ decreasing monotonically as scattering increases, then the sensitivity can be expressed as,
\begin{equation}
    \frac{d\gamma_{\text{OPP}}}{d\eta} = \frac{3}{2}m\frac{dm}{d\eta},\ \frac{d\gamma_{\text{TPP}}}{d\eta} = 3m^3\frac{dm}{d\eta}. 
\end{equation}
The crucial point is not merely the slope with respect to $m$, but the nonlinear compression of the state space. As $m$ decreases, the output states contract more rapidly, so equal changes in the physical parameter $\eta$ lead to disproportionately larger changes in the measured observable. Therefore, a given reduction in depolarizing channels produces a stronger contraction of the accessible correlation tensor and a stronger degradation of purity and entanglement. As shown in Fig.~\ref{fig:CPDE}, with input Bell state $\ket{\Psi^+}$, the simulated quantum metrics exhibit a markedly steeper decay with decreasing of parameter $m$.

\begin{figure*}[t]
    \centering
    \includegraphics[width=1\linewidth]{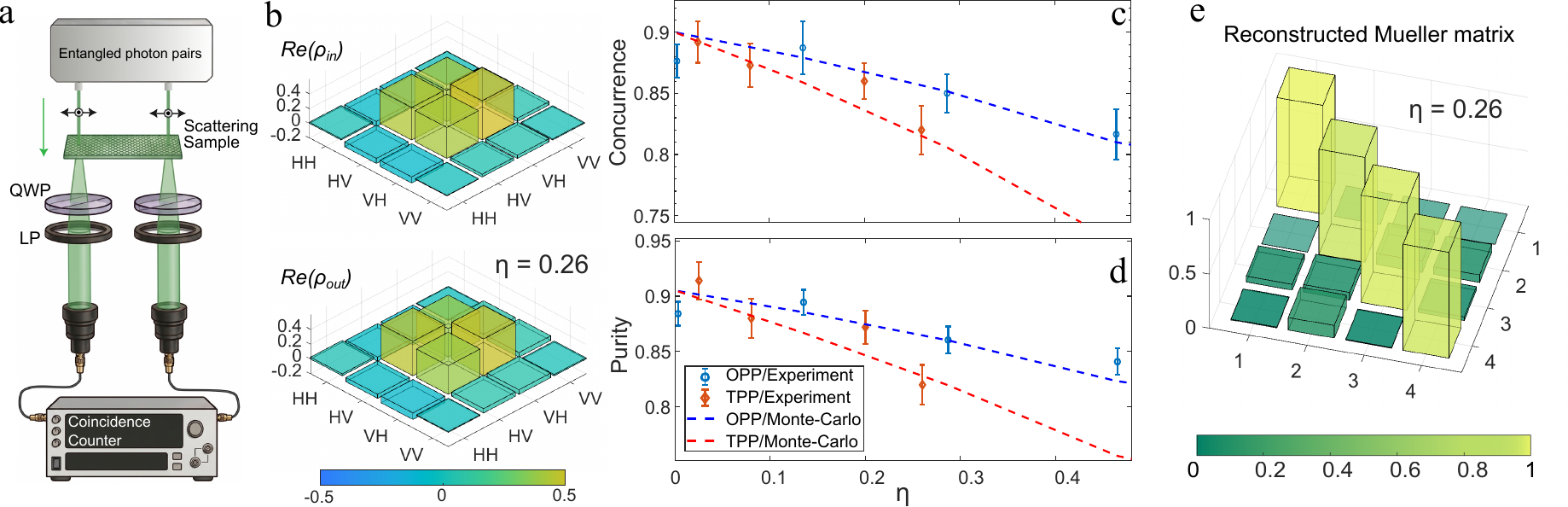}
    \caption{\textbf{a}.~ A simplified sketch of experimental setup for quantum two-photon polarimetry. %The photon pairs are generated from two ppKTP crystals in a Mach-Zehnder configuration. 
    Both photons of the entangled pairs are scattered and travel through the polarization state analyzer individually for quantum state tomography. \textbf{b}.~(Top) The real part of the density matrix of the prepared input state, concurrence of approx. 0.9. (Bottom) Output density matrix, real part, after sample with effective thickness $\eta=0.26$.  \textbf{c}.~Experimental concurrence change with effective thickness for OPP and TPP with Monte-Carlo simulation. \textbf{d}.~Experimental purity for OPP and TPP along with Monte-Carlo scattering simulation. \textbf{e}.~The reconstructed Mueller matrix of sample ($\eta =0.26$) from TPP experimental data through nonlinear fitting of Eqns.~(\ref{eq:CPTP}) and (\ref{eq:congurence}) by source density matrix and post-sample density matrix.}
    \label{fig:Experiment}
\end{figure*}

\emph{Experiment.} To verify these predictions, we perform experiments using polarization-entangled photon pairs generated via spontaneous parametric down-conversion in type-II periodically poled potassium titanyl phosphate (ppKTP) crystals in a polarization Mach-Zehnder interferometer~\cite{Lee:2016:Opt.Express, Fazili:2024}. Approaching the Bell state $\ket{\Psi^+}$, the experimentally prepared probing state (input) has a concurrence of 0.9. %, mainly due to residual compensation, distinguishability, and interferometric imbalance in the used SPDC source. 
The reconstructed input density matrix is shown in Fig.~\ref{fig:Experiment}b. This measured $\rho_{\text{in}}$ %, not an ideal Bell state,
is further used throughout the simulations and reconstructions, accounting for the finite concurrence in results interpretation. It mainly limits the dynamic range, without undermining the quadratic TPP response, which is consistent with earlier reported OPP with non-maximally entangled quantum states~\cite{Galvez:2025:Opt.Exp}. The photon pairs are directed toward tissue-mimicking phantom samples~\cite{Sieryi:2020:SPIE,Sieryi:2025:Biomed} with controlled effective thickness $\eta=d/l^* = [0.025, 0.08, 0.2, 0.26]$, where $d$ is the actual thickness (constant for all samples) and $l^*=1/\mu_s^\prime=\mu_s^{-1}\left(1-g\right)^{-1}$ is the transport mean free path with forward scattering anisotropy factor $g$, scattering coefficient $\mu_s$, and reduced scattering coefficient $\mu_s^\prime$~\cite{Alali:2015:JBO,Tuchin2015book}. Here, $\eta$ serves as an effective measure of scattering strength, with larger values corresponding to stronger depolarization, $m(\eta) \propto e^{-\alpha\eta}$, $\alpha>0$. In the weak-scattering limit, $m(\eta) \propto 1-\alpha\eta$.

Both photons pass through the sample with a small spatial separation that exceeds each photon width, ensuring a correlated interaction with the phantoms, while there is no interference between the photon pairs during the interaction. After transmission, each photon is analyzed independently using polarization state analyzers, and coincidence measurements over a complete tomographic basis are used to reconstruct the output density matrix~\cite{James:2001-52312:PRA}. A schematic setup is shown in Fig.~\ref{fig:Experiment}a, with details in Supplementary Material, Sec.~S3. For larger effective thickness, multiple scattering leads to significant angular broadening and wavefront distortion, making spatial mode overlap and coincidence alignment increasingly challenging, which limits the achievable count rate. The measurements are shot-noise limited.

To support the experimental observations, we perform Monte Carlo simulations of in-sample photon scattering following Ref.~\cite{lopushenko2024,meglinski2005,besaga2025LPR}. Each photon pair is propagated through the scattering medium by sampling stochastic trajectories characterized by the optical parameters of the sample, and being tracked using Bethe-Salpeter and radiative transfer formalisms~\cite{Salpeter1951PhysRev} (Supplementary Material Sec.~S4). Based on the Monte-Carlo simulated Jones matrices, polarization transformations are then represented by the Kraus operator $U_k$, describing the effective polarization evolution associated with a given scattering path. The output two-photon density matrix is then obtained as an ensemble average over $N \sim 10^6$ realizations via the CPTP map, Eq.~(\ref{eq:CPTP}). The corresponding Mueller matrix is reconstructed using Eq.~(\ref{eq:congurence}).

%Associated by the Monte-Carlo simulated in-sample photon scattering~\cite{lopushenko2024} (the TPP Monte-Carlo procedure is shown in Appendix.~\ref{sec: MC_TPP}) and the corresponding map presented in Kraus superoperator $U_k$, we simulate the density matrix evolution traveling through the depolarization properties of samples under statistical ensemble of $N\sim 10^6$ photon pairs, $k=1,..,N$, modeled via the CPTP map in Eq.~(\ref{eq:CPTP}). The corresponding Mueller matrices are reconstructed by the relation between $M$ and $U$ in Eq.~(\ref{eq:KMU_map}).

As shown in Figs.~\ref{fig:Experiment}c and~d, the experimentally measured TPP degradation agrees quantitatively with the simulated trends, exhibiting the predicted dependence on the depolarization strength. In the present weak-scattering regime, this quadratic response appears as an approximately two-fold enhancement of the first-order sensitivity relative to OPP, as reflected by a purity-slope ratio of $\sim 2.1$ (Supplementary Material, Sec. S3.7). For comparison, the corresponding OPP response is taken from our previous measurements performed using the same source of entangled photons and under similar conditions~\cite{besaga2025LPR}. This provides a direct experimental benchmark for the TPP results shown in Fig.~\ref{fig:Experiment}c,d %. This ensures a direct comparison between OPP and TPP 
without variations in experimental conditions and statistically consist with the two-fold enhancement through combined linear regression analysis~\cite{montgomery2021linear}. The contrast between these behaviors highlights the quadratic enhancement in sensitivity enabled by quantum two-photon polarimetry.

The isotropic depolarizing Mueller matrix is reconstructed from the experimentally measured input and output two-photon states via nonlinear fitting to a diagonal depolarization model. An example of the reconstructed matrix, corresponding to a sample with effective optical thickness of $\eta=0.26$, is shown in Fig.~\ref{fig:Experiment}e. The extracted parameters show a 97$\%$ agreement with %Monte Carlo 
simulations (quantified via normalized matrix similarity) of photon scattering in the sample, confirming the consistency between experiment and the underlying depolarization model. 

%The uniqueness of the reconstruction for anisotropic depolarizing channels with three independent parameters is established theoretically (see Supplementary~S2.2), and further verified through numerical simulations (Fig.~\ref{fig:REconstruction}), where a $201\times201$ Mueller-matrix polarimetric image is reconstructed from simulated two-photon data. The reconstruction error is below $10^{-8}$ and is limited solely by numerical fitting precision, thereby providing strong numerical evidence of uniqueness and confirming that the measured two-photon correlations provide sufficient constraints to fully determine the channel within this class.

%The stability and uniqueness of the reconstruction for anisotropic depolarizing channels with three independent parameters are further verified through numerical simulations (Fig.~\ref{fig:REconstruction}), demonstrating that the measured two-photon correlations provide sufficient constraints to fully determine the channel within this class.
\begin{figure}
    \centering
    \includegraphics[width=1\linewidth]{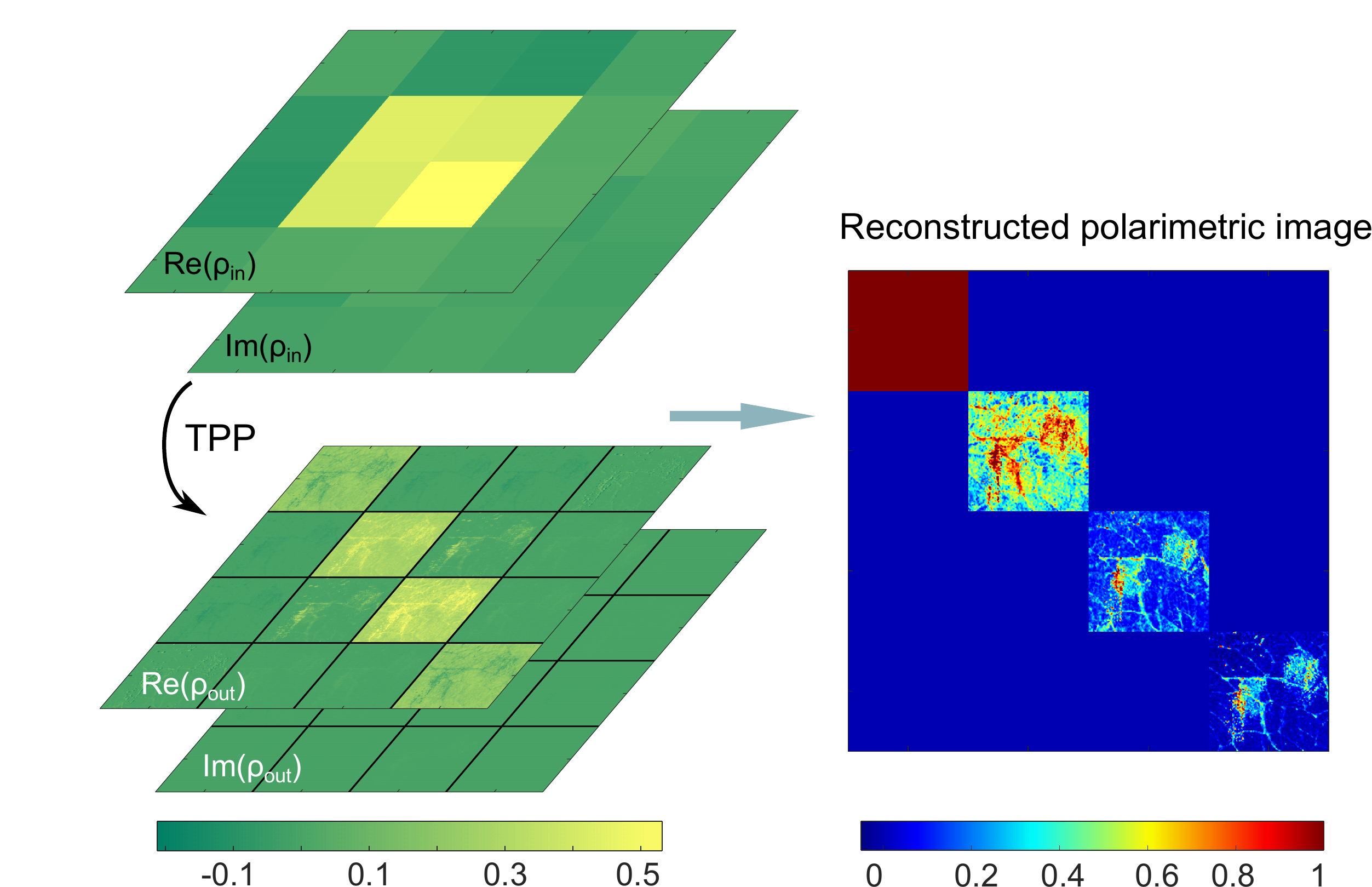}
    \caption{Reconstruction of a spatially varying depolarizing sample using quantum two-photon polarimetry. Left: measured real and imaginary components of the input sample-free two-photon density matrices $\rho_{\text{in}}$, obtained from quantum state tomography, together with the corresponding output density matrix, which is numerically simulated using Eq.~(\ref{eq:congurence}) with only the depolarization part from a Mueller matrix from Ref.~\cite{peyvasteh2021evolution}. Right: reconstructed polarimetric image with $201\times201$ pixels, where each pixel corresponds to the depolarization parameter obtained by fitting the input-output state transformation to a diagonal Mueller matrix model.}
    \label{fig:REconstruction}
\end{figure}

\emph{Reconstruction uniqueness.} The quadratic transformation in TPP marks a fundamental distinction from conventional quantum polarimetry: whereas standard measurements probe first-order polarization moments, TPP directly accesses second-order polarization correlations encoded in the two-photon state. This leads to qualitatively different measurement properties: in particular, the quadratic dependence on the Mueller matrix enhances sensitivity to depolarization, enabling detection of subtle polarization changes.

However, a direct consequence of this quadratic mapping is the emergence of reconstruction ambiguity for a generic $4\times 4$ Mueller matrix in the full 16-parameter space. Unlike the linear relation in one-photon polarimetry, the transformation in Eq.~(\ref{eq:congurence}) depends on the Mueller matrix through a congruence action. As a result, transformations of the form $M \to MO$ with $OK_{\text{in}}O^T = K_{\text{in}}$ leave the measurement invariant, leading to non-unique solutions. The set of such transformations defines the stabilizer group of the input correlation tensor~\cite{HornJohnson2012}, implying that a single input state is generally insufficient for complete reconstruction. This ambiguity can be resolved by employing multiple input states. For two inputs $K_{\mathrm{in}}^{(1)}$ and $K_{\mathrm{in}}^{(2)}$, the admissible transformations are restricted to $O \in \mathrm{Stab}(K_{\mathrm{in}}^{(1)}) \cap \mathrm{Stab}(K_{\mathrm{in}}^{(2)})$, which generically reduces to a discrete set, enabling unique reconstruction.
%up to a global sign. 
Physically, this reflects that TPQP probes only polarization correlations rather than the channel linearly, so distinct Mueller matrices can produce identical responses for a given input. Additional probe states break this symmetry by sampling complementary correlations, 
facilitating unambiguous reconstruction.
%restoring identifiability up to a global sign. 
Experimentally, we are limited to isotropic depolarizing samples, and reconstruction of general Mueller matrices via optimized input pairs is therefore validated numerically (see Supplementary Material, Sec.~S2).

For depolarizing channels as one of the reduced forms of Mueller matrix~\cite{chipman1995mueller,Agarwal:15,savenkov2012Springer} relevant to our experiment, where the Mueller matrix is diagonal,
%and low-dimensional, 
the reconstruction ambiguity is effectively removed. In this regime, TPP enables unique and minimal reconstruction from a single input state, demonstrating that practical reconstruction remains feasible despite the intrinsic nonlinearity. The uniqueness of the reconstruction for anisotropic depolarizing channels with three independent parameters is  further verified through numerical simulations (Fig.~\ref{fig:REconstruction}), where a $201\times201$ Mueller-matrix polarimetric image is reconstructed from simulated two-photon data. The reconstruction error is below $10^{-8}$ and is limited solely by fitting precision, thereby providing strong numerical evidence of uniqueness and confirming that the measured two-photon correlations provide sufficient constraints to fully determine the channel within this class. Therefore, for depolarizing channels, TPP retains the single-measurement reconstruction simplicity of OPP while providing quadratic sensitivity enhancement.

The theoretical framework developed here also establishes a direct connection between classical Mueller-matrix optics and quantum polarization channels. This connection enables polarization transformations in complex media to be analyzed using tools from quantum information theory.

\emph{Conclusion.} We have introduced and experimentally demonstrated two-photon quantum polarimetry as a novel approach for characterizing depolarizing channels in scattering media. By exploiting polarization-entangled photon pairs, the method probes second-order polarization correlations, leading to a quadratic dependence of the measured observables on the channel parameters. This enables direct reconstruction of the depolarizing Mueller matrix from experimentally measured input and output states. For anisotropic depolarizing channels with three independent parameters, we show that the two-photon measurement provides sufficient constraints for unique reconstruction within a minimal measurement framework without necessity to perform complete quantum process tomography~\cite{Mohseni:2008:PRA}. The experimentally reconstructed channel parameters are in agreement with theoretical predictions and Monte Carlo simulations, confirming both the accuracy and robustness of the method. More generally, our results highlight how multi-photon probing modifies the identifiability of polarization channels by accessing higher-order correlations beyond those available in first-order measurements. This demonstrates that two-photon probing is not an incremental sensitivity gain, but a distinct and fundamentally new sensing modality, suggesting broader applications in quantum sensing, imaging, and the study of complex optical environments. 
%This provides a new route toward efficient characterization of structured and scattering media, and suggests broader applications in quantum sensing, imaging, and the study of complex optical environments \textcolor{red}{with prospects for practical single-shot diagnostics}.

\begin{acknowledgments}
The authors thank Dr.~Evgeny~Lyubin for his assistance in classical characterization of samples. 
This work was supported by the Australian Research Council %(\url{https://doi.org/10.13039/501100000923}) 
Centre of Excellence for Transformative Meta-Optical Systems - TMOS (CE200100010), the German Research Foundation project Meta Active IRTG 2675 (437527638) and the EU Horizon Europe EIC Pathfinder Open Research and Innovation Programme (project OPTIPATH, No. 101185769). This work was partly funded by the German Federal Ministry of Research, Technology and Space (project ``QUANCER'', FKZ 13N16441). V.B. thanks to ProChance-career program of University of Jena. A.B. and I.M. acknowledge support from the EIC, European Cooperation in Science and Technology (COST) Action CA21159 %\textcolor{red}{\st{– ``Understanding light–biological surface interactions: possibility for new electronic materials and devices (}
(PhoBioS)
%\st{)'',}}
and %\textcolor{red}{\st{COST Action}} 
CA23125 %\textcolor{red}{\st{– ``The mETamaterial formalism approach to recognize cAncer (}
(TETRA)%\st{)''}}
.
\end{acknowledgments}

\emph{Data availability} --- The data that support the findings of this article are available
from the authors upon reasonable request.

%\nocite{*}

\bibliography{apssamp}% Produces the bibliography via BibTeX.

\end{document}